\documentclass[a4paper,aps,prl,superscriptaddress,twocolumn]{revtex4-1}
\usepackage{graphicx}
\usepackage{pstricks}
\usepackage{amsmath,bm}
\usepackage{pdfpages}

\makeatletter
\AtBeginDocument{\let\LS@rot\@undefined}
\makeatother

\newcommand{\ses}{FeSe$_{1-x}$S$_x$}
\newcommand{\tc}{$T_{\rm c}$}
\newcommand{\dw}{$d_{x^2-y^2}$}
\newcommand{\nef}{$N(E_{\rm F})$}

\begin{document}
\title{Two-dome Superconductivity in FeS Induced by a Lifshitz Transition}

\author{Makoto Shimizu} 
\affiliation{Department of Physics, Okayama University, Okayama 700-8530, Japan}
\author{Nayuta Takemori} 
\affiliation{Research Institute for Interdisciplinary Science, Okayama
  University, Okayama 700-8530, Japan}
\author{Daniel Guterding} 
\affiliation{Fachbereich Mathematik, Naturwissenschaften und Datenverarbeitung, 
Technische Hochschule Mittelhessen, Wilhelm-Leuschner-Stra{\ss}e 13, 
61169 Friedberg, Germany}
\author{Harald O. Jeschke}
\email{jeschke@okayama-u.ac.jp}
\affiliation{Research Institute for Interdisciplinary Science, Okayama
  University, Okayama 700-8530, Japan}

\begin{abstract}
Among iron chalcogenide superconductors, FeS can be viewed as a simple, highly compressed relative of FeSe without nematic phase and with weaker electronic correlations. Under pressure, however, the superconductivity of stoichiometric FeS disappears and reappears, forming two domes. We perform electronic structure and spin fluctuation theory calculations for tetragonal FeS in order to analyze the nature of the superconducting order parameter. In the random phase approximation we find a gap function with $d$-wave symmetry at ambient pressure, in agreement with several reports of a nodal superconducting order parameter in FeS. Our calculations show that, as a function of pressure, the superconducting pairing strength decreases until a Lifshitz transition takes place at 4.6 GPa. As a hole pocket with a large density of states appears at the Lifshitz transition, the gap symmetry is altered to sign-changing $s$-wave. At the same time the pairing strength is severely enhanced and increases up to a new maximum at 5.5 GPa. Therefore, our calculations naturally explain the occurrence of two superconducting domes in FeS.
\end{abstract}

% insert suggested PACS numbers in braces on next line
% \pacs{
%   71.20.-b, %Electron density of states and band structure of crystalline solids
%   71.20.-b, %DOS and band structure of crystalline solids
%   74.24.Ha, %Magnetic properties of superconductors
% }
\pacs{
74.20.Pq, %calculations in superconductivity of condensed matter
71.15.Mb, %Density Functional Theory, condensed matter
71.18.+y, %Fermi surface
74.70.Xa  %Chalcogenides, noncuprate superconductors
}

\maketitle

%Introduction section
%\section{Introduction}

% general
{\it Introduction.--} The structurally simplest class of iron-based
superconductors with its prime representative FeSe~\cite{Hsu2008} was
discovered in the same year as LaFeAsO~\cite{Kamihara2008}. FeSe has
been intensively studied due to its very large nematic
region~\cite{Boehmer2018}, its interesting
magnetism~\cite{Glasbrenner2015} and the complexity of its electronic
structure~\cite{Coldea2018}. Only in 2015 was it established that the
isostructural FeS is also a superconductor~\cite{Lai2015}. Even though
the replacement of Se by the smaller S appears to be a minor
structural modification, it soon became clear that FeSe and FeS behave
differently in several respects: The nematic region is absent in
FeS~\cite{Watson2015}, the electronic correlations appear to be
significantly smaller in FeS~\cite{Man2017}, and the upper critical
field is much smaller~\cite{Borg2016}. In fact, the possibility to
grow high quality mixed {\ses} structures has provided opportunities
to study the evolution of properties between FeSe and
FeS~\cite{Hosoi2016,Matsuura2017,Man2017,Sato2018,Hanaguri2018}.

% superconductivity

Superconductivity in FeS has been observed below $T_{\rm c}=
5$~K~\cite{Lai2015} with some variation due to sample
dependence~\cite{Kuhn2017}. Scanning tunneling spectroscopy points to
strong-coupling superconductivity~\cite{XYang2016}, and Hall
conductivities can be fitted with a two-band model~\cite{Lin2016}. The
symmetry of the superconducting gap in FeS has been the subject of
some debate. Using scanning tunneling spectroscopy, Yang {\it et
  al.}~\cite{XYang2016} conclude that the superconducting gap of FeS
is strongly anisotropic. Specific heat measurements~\cite{Xing2016}
and quasiparticle heat transport studies~\cite{Ying2016} point to a
nodal gap structure. However, muon spin rotation studies found fully
gapped behavior in
FeS~\cite{Holenstein2016,Kirschner2016}. Theoretically, a {\dw} order
parameter at ambient pressure has been obtained~\cite{YYang2016}.

% pressure

Pressure has been shown to suppress superconductivity in
FeS~\cite{Lai2016}. Surprisingly, however, Zhang {\it et al.} have
found that after the initial suppression, at a pressure of $P=5$~GPa
superconductivity reemerges, and a second superconducting dome is
formed, up to a pressure of $P=22.3$~GPa. Such double-dome
superconductivity is known to occur also in alkali iron
selenides~\cite{Sun2012} and in FeSe
intercalates~\cite{Izumi2015,Sun2018}. In fact, two superconducting
domes occur in nearly all classes of unconventional
superconductors~\cite{Das2016}.

In this Letter, we consider the structurally simple FeS as an
instructive example system for studying the origin of double-dome
superconductivity in iron-based materials. We show that at a pressure
of $P=4.6$~GPa, a Lifshitz transition occurs, adding a hole pocket to
the Fermi surface and boosting the density of states at the Fermi
level. Using spin fluctuation theory in the random phase
approximation, we show that pairing strength of the {\dw} order
parameter, which dominates within the low pressure dome, decreases
until a Lifshitz transition of the electronic structure takes
place. At the transition, the superconducting order parameter switches
to nodeless $s_\pm$, and the pairing strength grows significantly to a
new maximum. Our study highlights that even without a structural phase
transition, the pressure-induced changes in the electronic structure
trigger the reemergence of superconductivity in FeS.

% structure

{\it Structure.--} The metastable tetragonal structure of FeS
($P\,4/nmm$ space group) occurs as a mineral named
mackinawite~\cite{Lennie1995}. Single crystals can be synthesized by
hydrothermal synthesis~\cite{Lennie1995,Lai2015} and by
deintercalation of K$_x$Fe$_{2-y}$S$_2$~\cite{Borg2016}. We base our
study on the pressure series of tetragonal crystal structures
determined by Zhang {\it et al.}~\cite{Zhang2017}. In this study,
mackinawite is found to transform to the hexagonal troilite phase
($P\,\bar{6}2c$ space group) at high pressures, with a mixed region
extending from 5~GPa to 9.2~GPa. However, the high-pressure phase
diagram of FeS is complicated, and orthorhombic ($P\,nma$ space group)
and monoclinic ($P\,2_1a$ space group) phases have also been
described~\cite{Ehm2008,Lai2016}.

{\it Methods.--} We perform density functional theory calculations for
the tetragonal FeS structures within the full-potential local orbital
(FPLO)~\cite{Koepernik1999} basis, using the generalized gradient
approximation (GGA) exchange correlation functional~\cite{Perdew1996}
and fine $k$ meshes of $50\times 50\times 50$. We interpolate the
experimental crystal structure (as shown in Ref.~\cite{Supplement},
Fig.~S1), so that we can perform calculations employing fine pressure
steps of $0.1$~GPa. We construct ten-band tight-binding models using
the FPLO projective Wannier functions~\cite{Eschrig2009}, including
all Fe $3d$ states. We employ the unfolding method using point group
symmetries~\cite{Tomic2014} in order to obtain five-band tight-binding
models. We study superconductivity assuming a spin-fluctuation driven
pairing interaction within the multi-orbital Hubbard model and use the
formalism as detailed by Graser {\it et
  al.}~\cite{Graser2009,Supplement} and as implemented in
Refs.~\cite{Guterding2015a,Guterding2017}. We determine the
noninteracting susceptibilities on ${\bf q}$ meshes of $50\times
50\times 10$ points at all pressures, and we use about 5000 $k$ points
on the Fermi surface for solving the gap equation in three dimensions;
two-dimensional calculations are insufficient for FeS under pressure.

\begin{figure}[t]
\includegraphics[width=\linewidth]{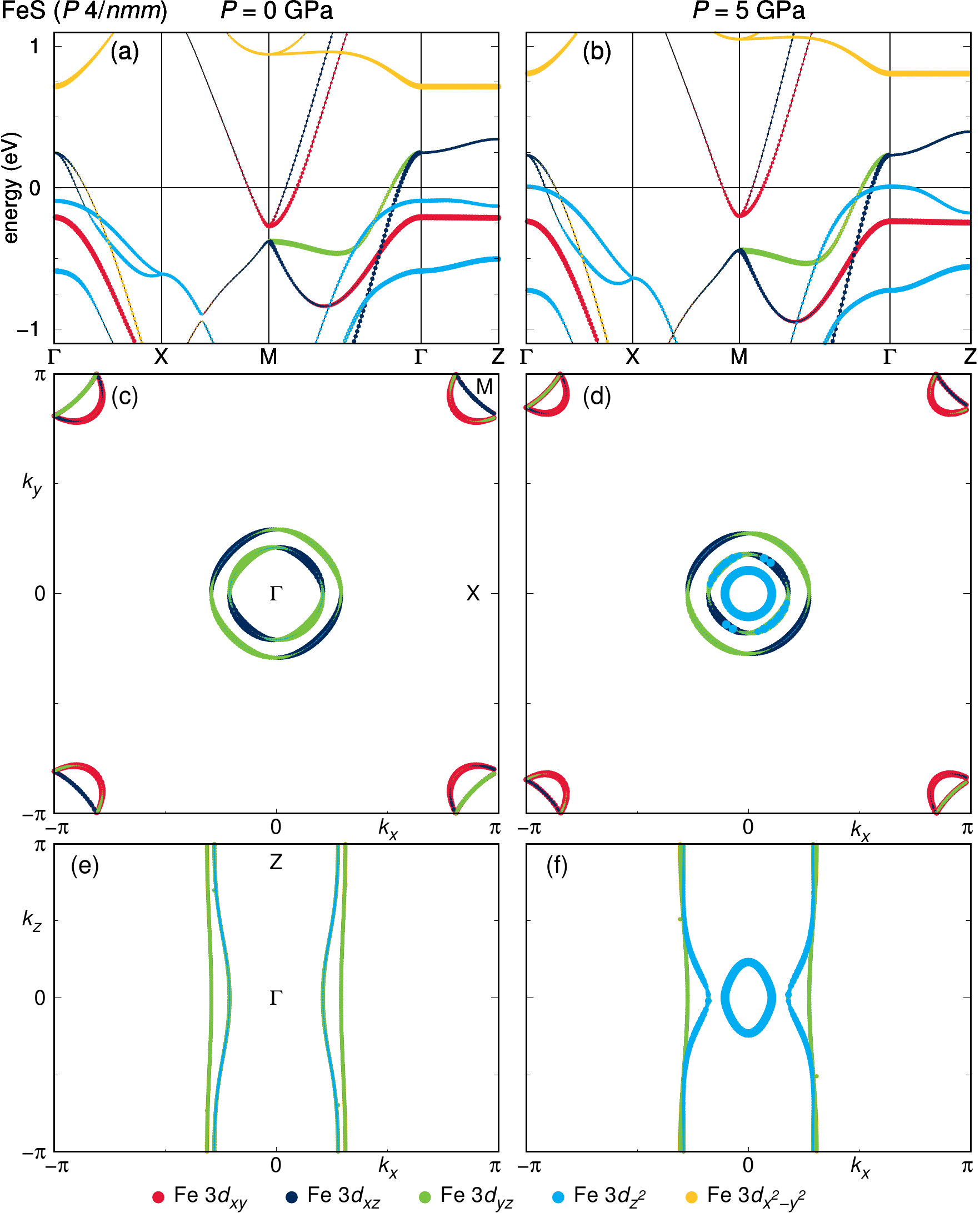}
\caption{Electronic structure of FeS at ambient pressure (left column)
  and at $P=5$~GPa (right column). Band structures (a), (b), Fermi
  surfaces in the $k_x-k_y$ plane at $k_z=0$ (c), (d) and Fermi
  surfaces in the $k_x-k_z$ plane (e), (f) are all colored with the
  orbital weights of the Fe $3d$ orbitals. A Lifshitz transition at
  $P=4.6$~GPa adds a hole Fermi surface pocket near
  $\Gamma$. }\label{fig:bandfs}
\end{figure}

  \begin{figure*}[t]
\includegraphics[width=\linewidth]{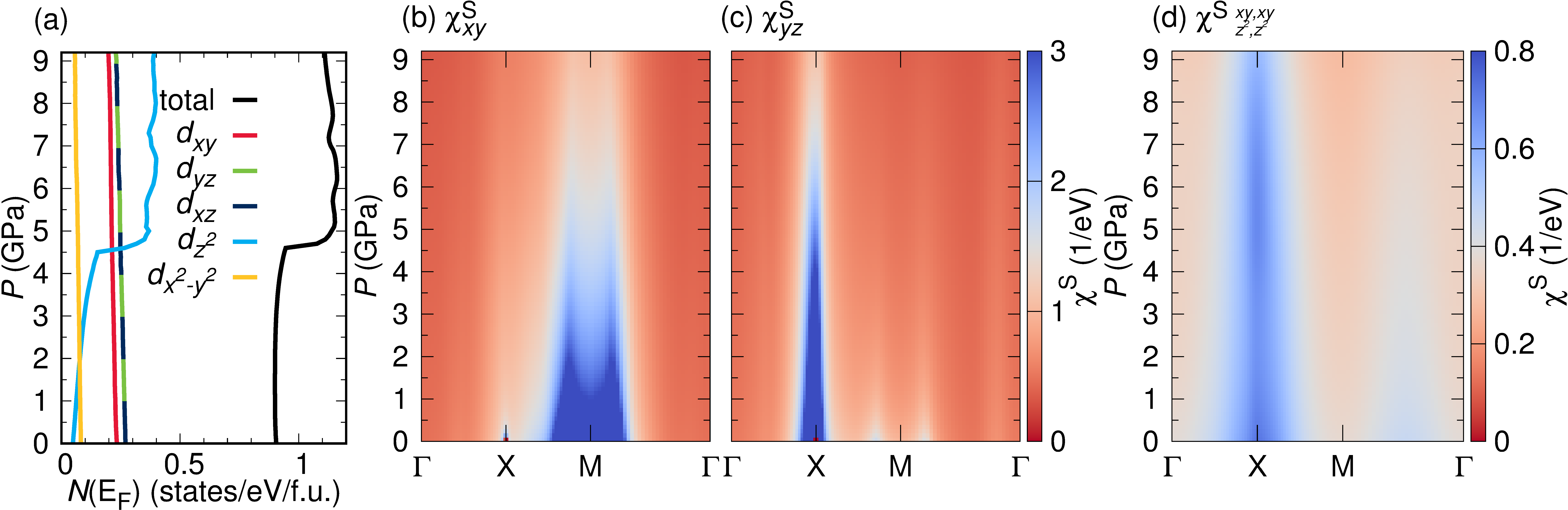}
\caption{Pressure dependence of (a) density of states at the Fermi
  level and (b), (c) diagonal elements and (d) off-diagonal elements
  of the spin susceptibility for tetragonal FeS.}\label{fig:nefsusc}
\end{figure*}

{\it Results.--} We first determine the electronic structure of
tetragonal FeS in small pressure intervals up to a pressure of
$P=9.2$~GPa. Fig.~\ref{fig:bandfs} shows bands and Fermi surfaces at
two representative pressures, $P=0$ and $P=5$~GPa. Our results at
ambient pressure are in good agreement with angle-resolved
photoemission~\cite{Miao2017,Reiss2017} and quantum oscillation
measurements~\cite{Terashima2016}. The fact that FeS is rather weakly
correlated~\cite{Reiss2017} makes the plain GGA calculations a good
starting point for our analysis of electronic structure and
superconductivity. After a very smooth pressure evolution of the
electronic structure, suddenly at $P=4.6$~GPa a Lifshitz transition
occurs and a hole pocket is added to the Fermi surface
(Fig.~\ref{fig:bandfs}~(b), (d), (f)). The reason for this event is
the fact that the bands with Fe $3d_{z^2}$ orbital character widen
more rapidly with pressure than the other iron bands. A careful analysis
of the relationship between geometrical parameters in the FeS
structure and its bands reveals that the $3d_{z^2}$ bands are
especially sensitive to the Fe-S-Fe angle, much more so than to the
Fe-S bond distance~\cite{Supplement}. As a consequence, the $3d_{z^2}$
contribution to the density of states at the Fermi level {\nef}
increases gradually below $P=4.6$~GPa before it rises by more than
100{\%} at the Lifshitz transition, as shown in
Fig.~\ref{fig:nefsusc}(a).

\begin{figure}[htb]
\includegraphics[width=\linewidth]{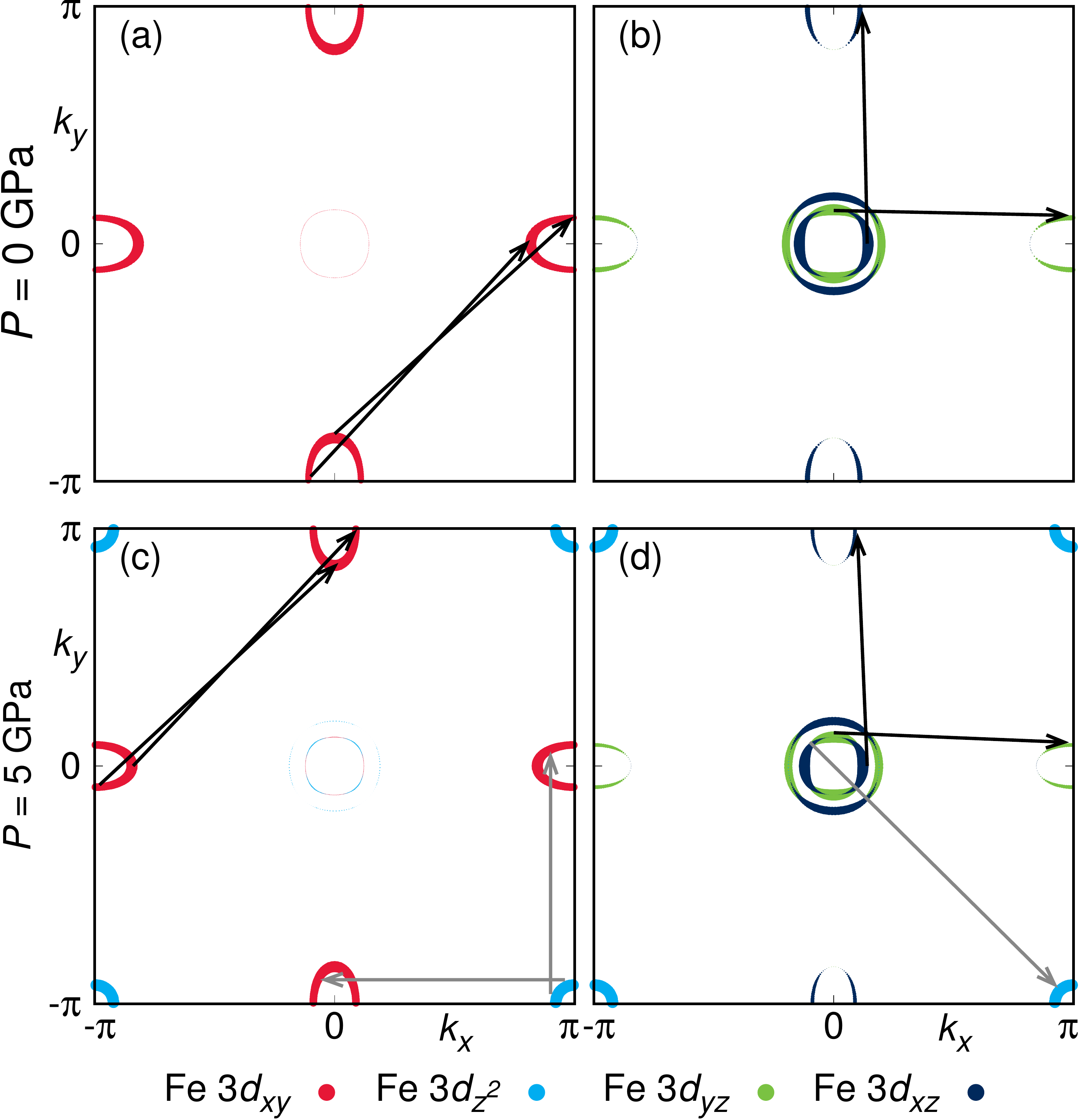}
\caption{ Fermi surfaces in the one-iron Brillouin zone at $k_z=0$ at
  ambient pressure and at $P=5$~GPa (after the Lifshitz
  transition). Black arrows indicate important intraorbital nesting
  vectors, gray arrows significant interorbital nesting
  vectors. }\label{fig:nesting}
\end{figure}

We now consider the superconductivity in FeS, assuming a
spin-fluctuation induced Cooper pairing. We use the random phase
approximation to calculate the spin susceptibility at all pressures
(for details see Ref.~\cite{Supplement}). In iron-based
superconductors, the pairing interaction is often dominated by
intraorbital nesting (see f.i.~Ref.~\cite{Guterding2015a}), and in
particular $\chi_{xy}^{\rm S}$ and $\chi_{yz}^{\rm S}$ (or
$\chi_{xz}^{\rm S}$), as shown in Figs.~\ref{fig:nefsusc}~(b) and (c),
respectively. These elements of the spin susceptibility are diagonal
in the four orbital indices, since we first investigate only
intraorbital contributions. The dominant peak in $\chi_{xy}^{\rm S}$
is near a nesting vector ${\bf q}=(\pi,\pi)$, in $\chi_{yz}^{\rm S}$
near ${\bf q}=(\pi,0)$. In fact, these nesting vectors can also be
extracted easily from a plot of the Fermi surface
(Fig.~\ref{fig:nesting}).
  
For repulsive interaction, a peak in the spin susceptibility at vector
$\bf{q}$ induces a sign change of the superconducting gap between
Fermi surface pockets connected by $\bf{q}$. From the spin
susceptibility, it is clear that the electronic structure of FeS leads
to the competition between different order parameters, which is
typical for iron-based superconductors. The peak at ${\bf q}=(\pi,0)$
in $\chi_{yz}^{\rm S}$ favors a sign change between hole cylinders
around $\Gamma$ and electron cylinders at $X$ and $Y$, {\it i.e.}~a
type of sign-changing $s$-wave order parameter, where the gap has the
same sign on all electron pockets. On the other hand, the strong ${\bf
  q}=(\pi,\pi)$ peak in $\chi_{xy}^{\rm S}$ favors a sign change
between the electron pockets; this is most easily fulfilled by a {\dw}
order parameter. As a compromise between these two possibilities, a
nodal sign-changing $s$-wave order parameter sometimes occurs (see
f.i.~Ref.~\cite{Guterding2015a}).

\begin{figure*}[t]
\includegraphics[width=\linewidth]{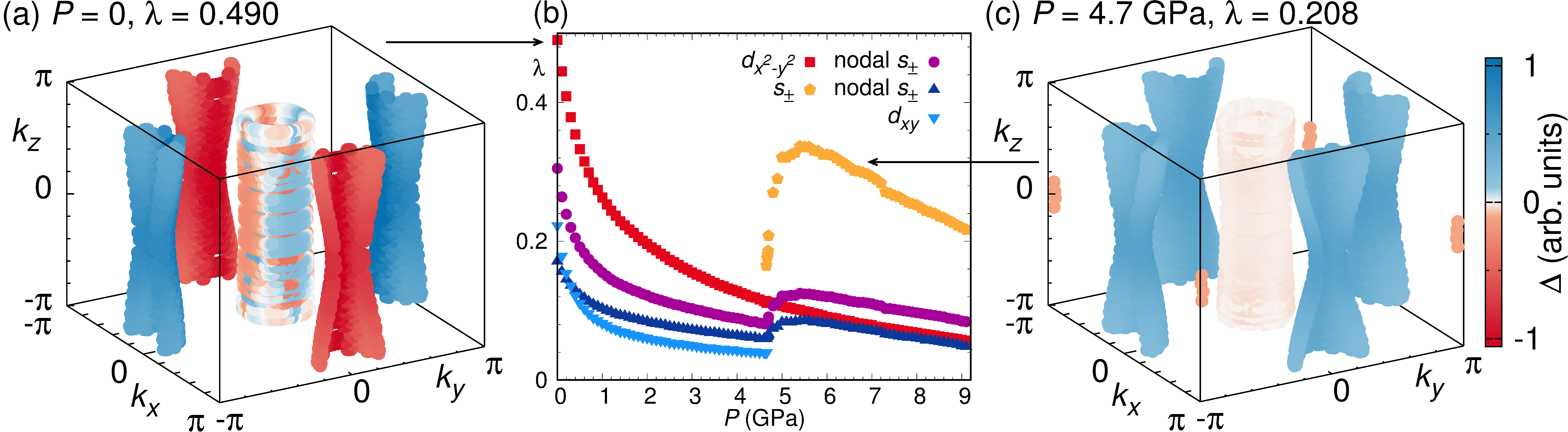}
\caption{Leading gap functions for FeS at (a) ambient pressure and (c)
  $P=4.7$~GPa. The three-dimensional Fermi surfaces are plotted in the
  one-iron Brillouin zone. (b) Leading eigenvalues $\lambda$ of the
  linearized gap equation as function of pressure. Up to $P=4.6$~GPa,
  the {\dw} order parameter dominates, but its eigenvalue decreases,
  marking the right half of the first superconducting dome. At the
  Lifshitz transition pressure $P=4.6$~GPa, an $s_\pm$ solution takes
  over, with its eigenvalue forming a second superconducting
  dome. }\label{fig:pairing}
\end{figure*}

Because of the increased bandwidth and smaller density of states at
the Fermi level, the spin susceptibility generally decreases with
increasing pressure (Fig.~\ref{fig:nefsusc}(b) and (c)). Therefore, we
find a general decline of pairing strength with increasing pressure
(Fig.~\ref{fig:pairing}(b)). From a quantitative solution of the
superconducting gap equation, we find (Fig.~\ref{fig:pairing}~(a))
that at $P=0$, the {\dw } solution wins, while several $s_{\pm}$
solutions are in competition, but subleading
(Fig.~\ref{fig:pairing}~(b)). This result is in agreement with
Ref.~\cite{YYang2016}. As a function of pressure, the eigenvalues of
the gap equation are suppressed rapidly. Initially, no change in the
symmetry of the superconducting gap is found. This corresponds well to
the first superconducting dome that was observed
experimentally~\cite{Zhang2017}.

However, very close to the Lifshitz transition, the nature of the
superconducting order parameter changes dramatically. At $P=4.65$~GPa,
a new sign-changing $s$ type order parameter appears
(Fig.~\ref{fig:pairing}~(c)) and becomes the dominating solution up to
the highest pressure $P=9.2$~GPa, at which the tetragonal phase is
completely replaced by the hexagonal phase of FeS. The eigenvalue of
the gap equation increases rapidly for this solution, in very good
agreement with the experiment, up to a maximum at $P=5.4$~GPa. Thus, our
calculation provides clear evidence for the existence of two dome
superconductivity in FeS under pressure. At the present level of theory we cannot compare superconducting and nonsuperconducting ground states which means that we have a strong suppression of the superconducting pairing strength but cannot capture a $T_c=0$ pressure interval.

Note that the eigenvalue of the nodal $s_\pm$ solution is also
enhanced at the Lifshitz transition, while the eigenvalue of the {\dw}
solution is not affected at all (Fig.~\ref{fig:pairing}(b)). This is
the case, because the symmetry-required nodes of the {\dw} solution
are located exactly where the $d_{z^2}$ hole pocket
emerges. Therefore, it is naturally excluded from the {\dw} solution.

As we have not studied superconductivity in the hexagonal phase, which
is presumably of nonmagnetic, BCS origin, we cannot complete the
second superconducting dome at the higher pressures investigated
experimentally. Predicting the internal coordinates for the
$P\,\bar{6}2c$ space group FeS structures at high pressures based on
experimental lattice parameters and analyzing the superconducting
mechanism is an interesting endeavour which is beyond the scope of the
present study.

{\it Discussion.--} So far, we have demonstrated two important effects
that occur in pressurized FeS without any structural discontinuity: a
Lifshitz transition which creates a hole pocket and significant Fe
$3d_{z^2}$ weight at the Fermi level, and a change of superconducting
order parameter from $d$ to sign-changing $s$-wave, which occurs at
almost exactly the same pressure. The important question of the
connection between the two events remains to be answered.

While the noninteracting diagonal susceptibility $\chi^0_{d_{z^2}}$
acquires some weak maximum near ${\bf q}=0$ (see
Ref.~\cite{Supplement}), the diagonal spin susceptibility
$\chi^S_{d_{z^2}}$ is nearly featureless and does not help to explain
any change in superconducting order parameter. Note, that the hole
Fermi surface around $M$ in the unfolded one-iron Brillouin zone (see
Fig.~\ref{fig:pairing}~(c)) is of different nature from the $\gamma$
Fermi surface feature at $M$ described in Ref.~\cite{Kuroki2009}; in
their case, electron doping populates a pocket of $d_{xy}$ orbital
character (in the local coordinates chosen for the present analysis),
and the pocket can contribute to pairing via the ${\bf q}=(\pi,\pi)$
peak in $\chi^S_{xy}$.

Our case highlights the importance of the interaction terms
proportional to $U'$, $J$ and $J'$: They mediate participation of the
$d_{z^2}$ orbital in the pairing via the off-diagonal components of
the spin susceptibility, such as ${\chi^S}_{aa}^{bb}$,
${\chi^S}_{ab}^{ba}$ and ${\chi^S}_{ab}^{ab}$ with $a=d_{z^2}$ and
$b=d_{xy}$ (or, in principle, also $b=d_{xz / yz}$) (see
Fig.~\ref{fig:nefsusc}(d) and Ref.~\cite{Supplement}), which are
significantly peaked at ${\bf q}=(\pi,0)$. The interorbital nesting
between $d_{z^2}$ and $d_{xz}$ is much weaker and does not contribute
significantly to the pairing. Figure~\ref{fig:nesting} shows the
relevant intra- and interorbital nesting vectors before and after the
Lifshitz transition. This figure indeed confirms that there is
considerable interorbital nesting between the $d_{xy}$ and $d_{z^2}$
orbitals.

Although Figs.~\ref{fig:nesting}(c) and (d) show that there is only a
small pocket of $d_{z^2}$ character, its strong influence on the
pairing interaction is explained by the extremely large density of
states at the Fermi level in this orbital after the Lifshitz
transition (Fig.~\ref{fig:nefsusc}(a)).

Finally, we also comment on the negligible gap size on the central
hole pockets in Fig.~\ref{fig:pairing}(c). The intraorbital spin
susceptibility of the $d_{xz/yz}$ orbitals, which is peaked at $X$,
should lead to a sign change between the electron pockets and the
central hole pockets as discussed before, with negative sign on the
central hole pockets. However, the interorbital spin susceptibility
between $d_{z^2}$ and $d_{xz/yz}$, which is peaked at $M$, should lead
to a sign change between the emergent hole pocket and the central hole
pockets, with positive sign on the central hole pockets. Therefore,
these interactions are frustrated. As a compromise, the gap on the
central hole pockets remains close to zero.

Since our analysis highlights the importance of interorbital
interactions, one could expect that superconductivity breaks down once
interorbital Coulomb interaction, Hund's rule coupling and
pair-hopping term are neglected. We corroborate the significance of
the interorbital interaction terms by solving the gap equation at
finite intraorbital Coulomb interaction $U=1.9$~eV with other
interactions set to zero ($U'=J=J'=0$). The order parameter we obtain
in this case is nodeless $s_\pm$, but, more importantly, the
associated pairing eigenvalue is close to zero, {\it
  i.e.}~superconductivity vanishes without interorbital interactions.

It would be very interesting to probe the Lifshitz transition by performing quantum oscillation experiments in FeS at pressures around 5~GPa; also, the predicted superconducting order parameter change could be observed in low-temperature specific heat measurements under pressure.

{\it Conclusion.--} We investigated the superconducting order
parameter of tetragonal FeS using a combination of density functional
theory calculations and spin fluctuation theory for the multi-orbital
Hubbard model. We showed that a Lifshitz transition occurs in FeS at a
pressure of about $P=4.6\,\mathrm{GPa}$, which changes the
superconducting order parameter from {\dw} to a sign-changing
$s$-wave, with significantly enhanced pairing strength right after the
Lifshitz transition due to enhanced density of states at the Fermi
level. While superconducting pairing within the first dome is
dominated by intraorbital nesting of $d_{xy}$ states, the second dome
features unusual interorbital nesting between $d_{xy}$ and $d_{z^2}$
states. In conclusion, our calculations explain the recently found
double-dome superconductivity in FeS.

\begin{acknowledgments}
  We acknowledge fruitful discussions with Seiichiro Onari, Hiroaki Ikeda and
  Kazuhiko Kuroki. N.T. is supported by JSPS KAKENHI Grant
  No. 16H07447. Part of the computations was carried out at the
  Supercomputer Center at the Institute for Solid State Physics, the
  University of Tokyo. 
\end{acknowledgments}

%\end{document} 

\clearpage


\section*{Supplementary material}

# Two-dome Superconductivity in FeS Induced by a Lifshitz Transition
– Supplemental Material –

Makoto Shimizu,[1] Nayuta Takemori,[2] Daniel Guterding,[3] and Harald O. Jeschke[2, *]

[1]*Department of Physics, Okayama University, Okayama 700-8530, Japan*
[2]*Research Institute for Interdisciplinary Science, Okayama University, Okayama 700-8530, Japan*
[3]*Fachbereich Mathematik, Naturwissenschaften und Datenverarbeitung,
Technische Hochschule Mittelhessen, Wilhelm-Leuschner-Straße 13, 61169 Friedberg, Germany*

## A. Structure

We use the crystal structures for tetragonal FeS ($P4/nmm$ space group) as given in Ref. [1]. The structure is defined by $a$ and $c$ lattice parameters and the S height $h_\text{S}$ above the iron plane as shown in Figure S1 (a), (b) and (c), respectively. Note that the slightly decreasing height of S above the Fe plane as function of pressure (Figure S1 (c)) translates into a slightly increasing S $z$ fractional coordinate. Relaxation of this position using density functional theory within GGA reverses this trend and is thus unreliable for FeS.

Concerning the reliability of the interpolation, we observe that the experimental data points in particular for the sulphur height and to a lesser extent also for the lattice parameters show some scatter. This is, intentionally, not reproduced by our interpolation. For the $c$ lattice parameter, for example, the experimental value at 4.5 GPa is 4.81 Å while our interpolation is 4.78 Å. This is justified by the experimental error bars. Concerning the sulphur height, the overall variation of only 0.06 Å in the entire pressure range is rather small. At 4.5 GPa, our interpolated values differs by 0.01 Å from the experimental data point which is only a modest deviation. Determination of the internal coordinates is also experimentally somewhat less straight-forward than the dermination of the lattice constants. Please note that a less monotonous interpolation than the one we chose would lead to unphysically strong variations in the elastic constants.

## B. Origin of the Lifshitz transition

We investigate which structural change in FeS under pressure is most important for the occurrence of the Lifshitz transition at $P = 4.6$ GPa. Bond length $d_\text{Fe-S}$ and $\delta \equiv \angle(\text{Fe-S-Fe})$ in tetragonal FeS are related to lattice constants $a$ and $c$ and sulfur height $h_\text{S}$ via

$$
\begin{aligned}
d_\text{Fe-S} &= \sqrt{\frac{a^2}{4} + h_\text{S}^2} \\
\sin\frac{\delta}{2} &= \frac{a}{2\sqrt{2}\, d_\text{Fe-S}} = \frac{a}{\sqrt{2a^2 + 8h_\text{S}^2}}
\end{aligned} \quad \text{(S1)}
$$

---

[*] jeschke@okayama-u.ac.jp

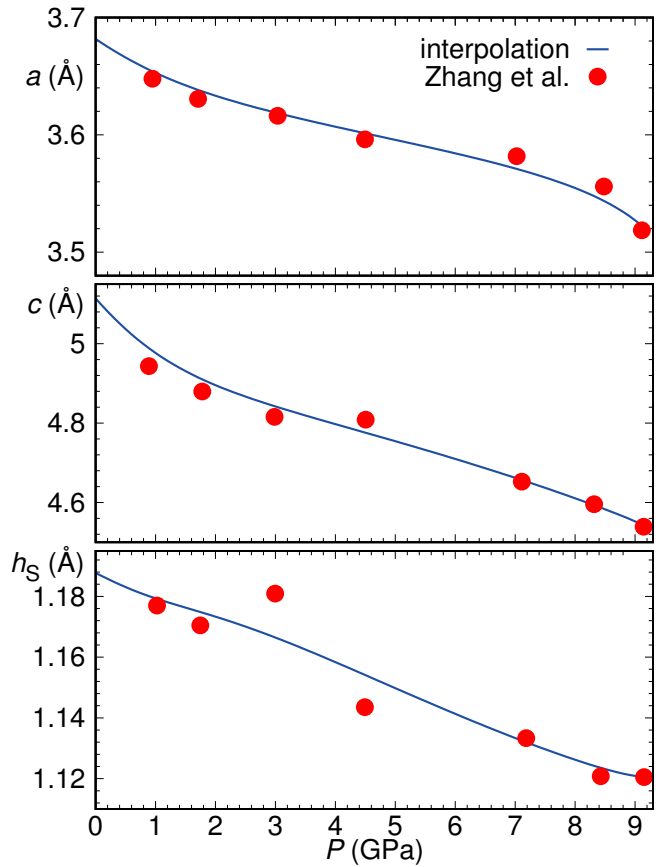

FIG. S1. Experimental crystal structures as measured by Zhang et al. [1] (symbols) together with Bézier interpolation (lines). (a), (b) are the tetragonal lattice parameters, and (c) is the height $h_\text{S}$ of S above the Fe plane.

Figure S2 shows the $d_\text{Fe-S}$ and $\delta$ calculated for the interpolated series of structures as function of pressure. We now investigate the sensitivity of the electronic structure and in particular the unoccupied band with Fe $3d_{z^2}$ orbital character near the Fermi level to the two relevant structural parameters separately. We focus on structures near $P = 4.6$ GPa. Upon fixing Fe-S distances $d_\text{Fe-S}$ and Fe-S-Fe angles $\delta$ to the values indicated by the dashed lines in Figure S2, we calculate $a$ lattice parameter and

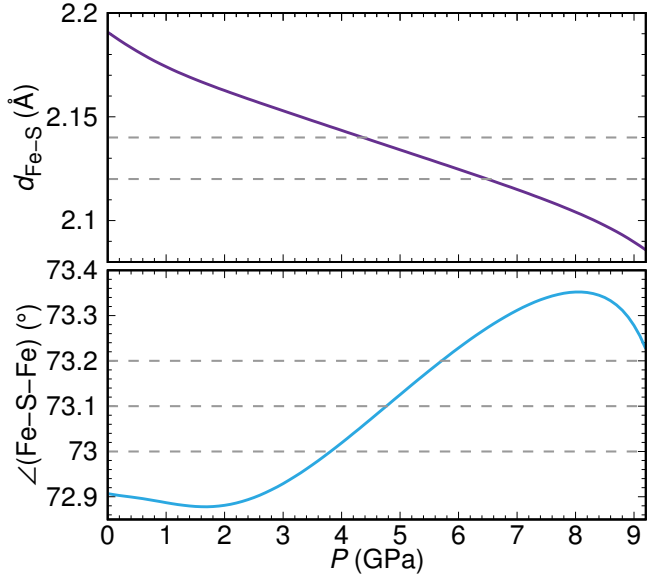

FIG. S2. Variation of (a) Fe-S distance and (b) Fe-S-Fe angle as function of pressure. Dashed lines indicate the bond distances and angles chosen for the plots in Fig. S3.

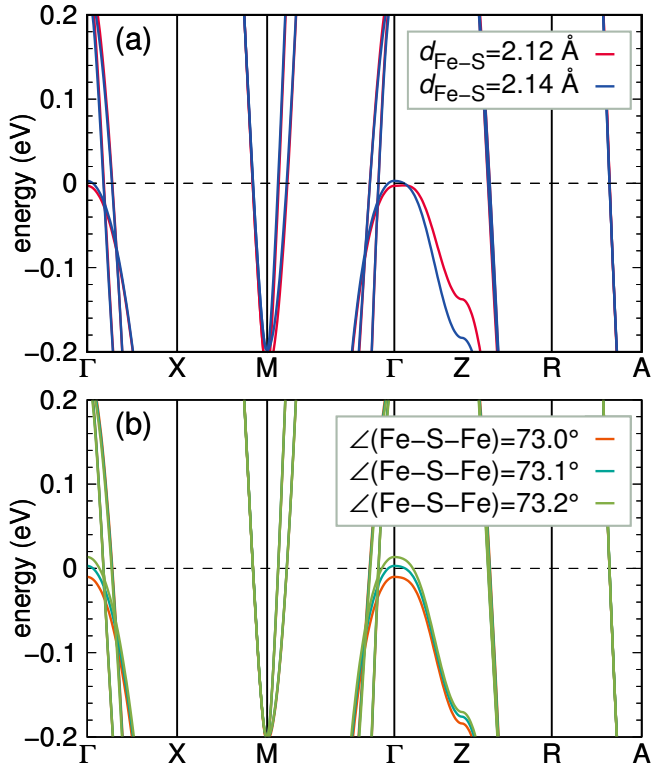

FIG. S3. Change of band structure of FeS for changes in (a) Fe-S distance and (b) Fe-S-Fe angle.

S $z$ coordinate $z_S$ by

$$a = 2\sqrt{2}d_{\text{Fe-S}}\sin\frac{\delta}{2}$$
$$z_S = \frac{1}{c}\sqrt{d_{\text{Fe-S}}{}^2 - \frac{a^2}{4}} \quad \text{(S2)}$$

Figure S3 shows the result. A very small change in band structure results from the substantial Fe-S bond length change corresponding to a pressure increase of roughly 2 GPa (Figure S3 (a)) once the Fe-S-Fe angle is kept constant; on the other hand, the effect of a pure angle change as is actually relevant in a 2 GPa window leads to a substantial change in band structure (Figure S3 (b)) even if the Fe-S bond lengths are fixed. Thus, the Lifshitz transition can be considered mainly an effect of the deformation of the FeS$_4$ tetrahedron rather than its pressure induced volume reduction.

### C. Spin fluctuation formalism

We follow Graser *et al.* [2] in considering the multi-orbital Hubbard model

$$\begin{aligned}H = H_0 &+ U\sum_{i,l}n_{il\uparrow}n_{il\downarrow} \\ &+ \frac{U'}{2}\sum_{i,s,p\neq s}n_{is}n_{ip} - \frac{J}{2}\sum_{i,s,p\neq s}\boldsymbol{S}_{is}\cdot\boldsymbol{S}_{ip} \\ &+ \frac{J'}{2}\sum_{i,s,p\neq s,\sigma}c^\dagger_{is\sigma}c^\dagger_{is\bar{\sigma}}c_{ip\bar{\sigma}}c_{ip\sigma}\end{aligned} \quad \text{(S3)}$$

where $c^\dagger_{is\sigma}$ ($c_{is\sigma}$) are Fermionic creation (annihilation) operators, $\boldsymbol{S}_{is}$ is the spin operator, $n_{is\sigma} = c^\dagger_{is\sigma}c_{is\sigma}$, $U$ denotes the intraorbital Coulomb repulsion, $U'$ denotes the interorbital Coulomb repulsion, $J$ denotes the Hund's rule coupling and $J'$ denotes the pair-hopping term. The tight binding part of the Hamiltonian is

$$H_0 = -\sum_{i,j}t^{sp}_{ij}c^\dagger_{is\sigma}c_{jp\sigma} \quad \text{(S4)}$$

where $t_{ij}$ denotes the transfer integral between sites $i$ and $j$, $s$ and $p$ are the orbital indices, and $\sigma$ denotes the spin index. The five band tight binding Hamiltonian is obtained using projective Wannier functions [3] and unfolding [4]. We first calculate the static noninteracting susceptibility

$$\begin{aligned}\chi^{pq}_{st}(\boldsymbol{q}) = -\sum_{\boldsymbol{k},l,m}&a^{p*}_l(\boldsymbol{k})a^t_l(\boldsymbol{k})a^{s*}_m(\boldsymbol{k}+\boldsymbol{q})a^q_m(\boldsymbol{k}+\boldsymbol{q}) \\ &\times\frac{n_F(E_l(\boldsymbol{k})) - n_F(E_m(\boldsymbol{k}+\boldsymbol{q}))}{E_l(\boldsymbol{k}) - E_m(\boldsymbol{k}+\boldsymbol{q})}\end{aligned} \quad \text{(S5)}$$

where $E_l(\boldsymbol{k})$ is the energy value determined by the band index $l$ and the wave vector $\boldsymbol{k}$, and $n_F(E)$ is the Fermi

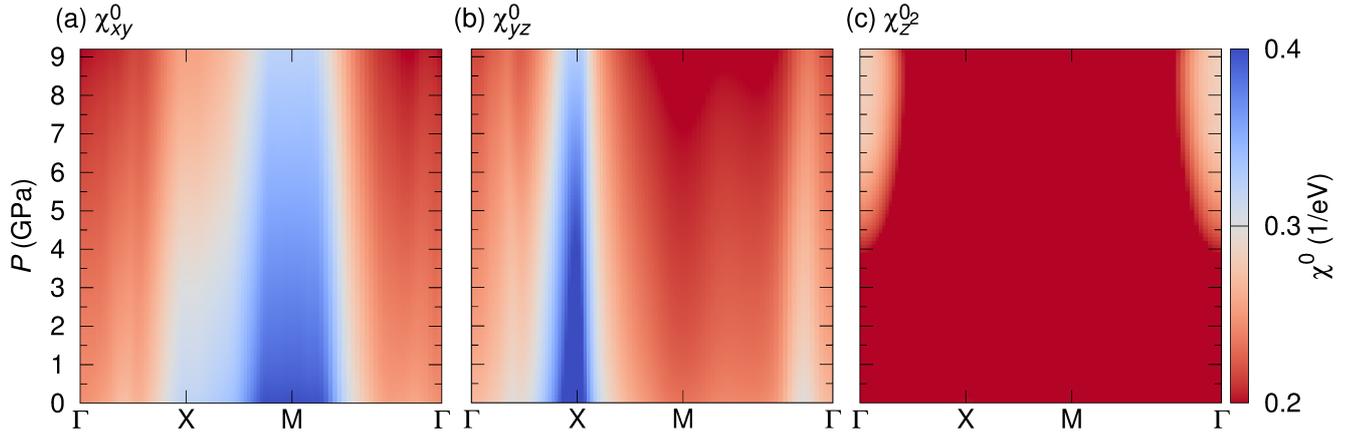

FIG. S4. Pressure dependence of the static noninteracting susceptibility for tetragonal FeS.

distribution function. $a_m^s$ is the matrix element of eigenvectors resulting from diagonalization of tight-binding Hamiltonian $H_0$. Within the framework of the random phase approximation (RPA) the charge and spin susceptibilities can be calculated from the noninteracting susceptibility

$$\left[(\chi_c^{RPA})_{st}^{pq}\right]^{-1} = [\chi_{st}^{pq}]^{-1} + (U_c)_{st}^{pq}$$
$$\left[(\chi_s^{RPA})_{st}^{pq}\right]^{-1} = [\chi_{st}^{pq}]^{-1} - (U_s)_{st}^{pq}$$
(S6)

where the nonzero components of the interaction tensors for the multi-orbital Hubbard model are given by [2]

$$(U_c)_{aa}^{aa} = U \qquad (U_c)_{bb}^{aa} = 2U',$$
$$(U_c)_{ab}^{ab} = \frac{3}{4}J - U' \qquad (U_c)_{ab}^{ba} = J'$$
$$(U_s)_{aa}^{aa} = U \qquad (U_s)_{bb}^{aa} = \frac{1}{2}J$$
$$(U_s)_{ab}^{ab} = \frac{1}{4}J + U' \qquad (U_s)_{ab}^{ba} = J', \qquad \text{(S7)}$$

which enables us to calculate the two-electron pairing vertex. For the interaction parameters, we find that the choice $U = 1.90$ eV, $U' = U/2$, $J = U/4$ and $J' = U/4$ takes us near the instability for all structures.

*Pairing calculations.–* The superconducting pairing vertex in the singlet channel is given by

$$\Gamma_{st}^{pq}(\mathbf{k}, \mathbf{k}')$$
$$= \left[\frac{3}{2} U_s \chi_s^{RPA}(\mathbf{k} - \mathbf{k}') U_s + \frac{1}{2} U_s \right.$$
$$\left. - \frac{1}{2} U_c \chi_c^{RPA}(\mathbf{k} - \mathbf{k}') U_c + \frac{1}{2} U_c \right]_{ps}^{tq}.$$
(S8)

The vertex in the orbital space description can be projected onto band space using the eigenvector resulting from diagonalization of the tight-binding Hamiltonian,

$$\Gamma_{ij}(\mathbf{k}, \mathbf{k}')$$
$$= \sum_{s,t,p,q} a_i^{t*}(-\mathbf{k}) a_i^{s*}(\mathbf{k}) \text{Re}\left[\Gamma_{st}^{pq}(\mathbf{k}, \mathbf{k}')\right] a_j^p(\mathbf{k}') a_j^q(-\mathbf{k}').$$
(S9)

Using the vertex $\Gamma_{ij}(\mathbf{k}, \mathbf{k}')$, we solve the gap equation

$$-\sum_j \oint_{C_j} \frac{dk'_\parallel}{2\pi} \frac{1}{4\pi v_F(\mathbf{k}')} \left[\Gamma_{ij}(\mathbf{k}, \mathbf{k}') + \Gamma_{ij}(\mathbf{k}, -\mathbf{k}')\right] g_j(\mathbf{k}')$$
$$= \lambda_i g_i(\mathbf{k})$$
(S10)

where $\lambda_i$ denotes the pairing eigenvalue and $g_i(\mathbf{k})$ is the gap function.

### D. Noninteracting susceptibility

Figure S4 shows the static noninteracting susceptibilities of FeS for the orbitals that have significant weight at the Fermi level, (a) $d_{xy}$, (b) $d_{yz}$ (which is by symmetry equivalent to $d_{xz}$ in tetragonal FeS), and (c) $d_{z^2}$. The latter only acquires features near $\Gamma$ after the Lifshitz transition because the orbital weight is concentrated on one hole pocket.

### E. One-iron Fermi surfaces of FeS

Figure S5 gives the orbital character of FeS Fermi surfaces at two $k_z$ values, $k_z = 0$ and $k_z = \pi$ at ambient pressure and at $P = 5.5$ GPa, after the Lifshitz transition. At $P = 0$, only little $3d_{z^2}$ weight is present in the hole pockets near $\Gamma$. The new hole pocket around $M$ has mostly $3d_{z^2}$ character. Comparison of the sizes of the electron pockets around $(\pi, 0, k_z)$ and $(0, \pi, k_z)$ between

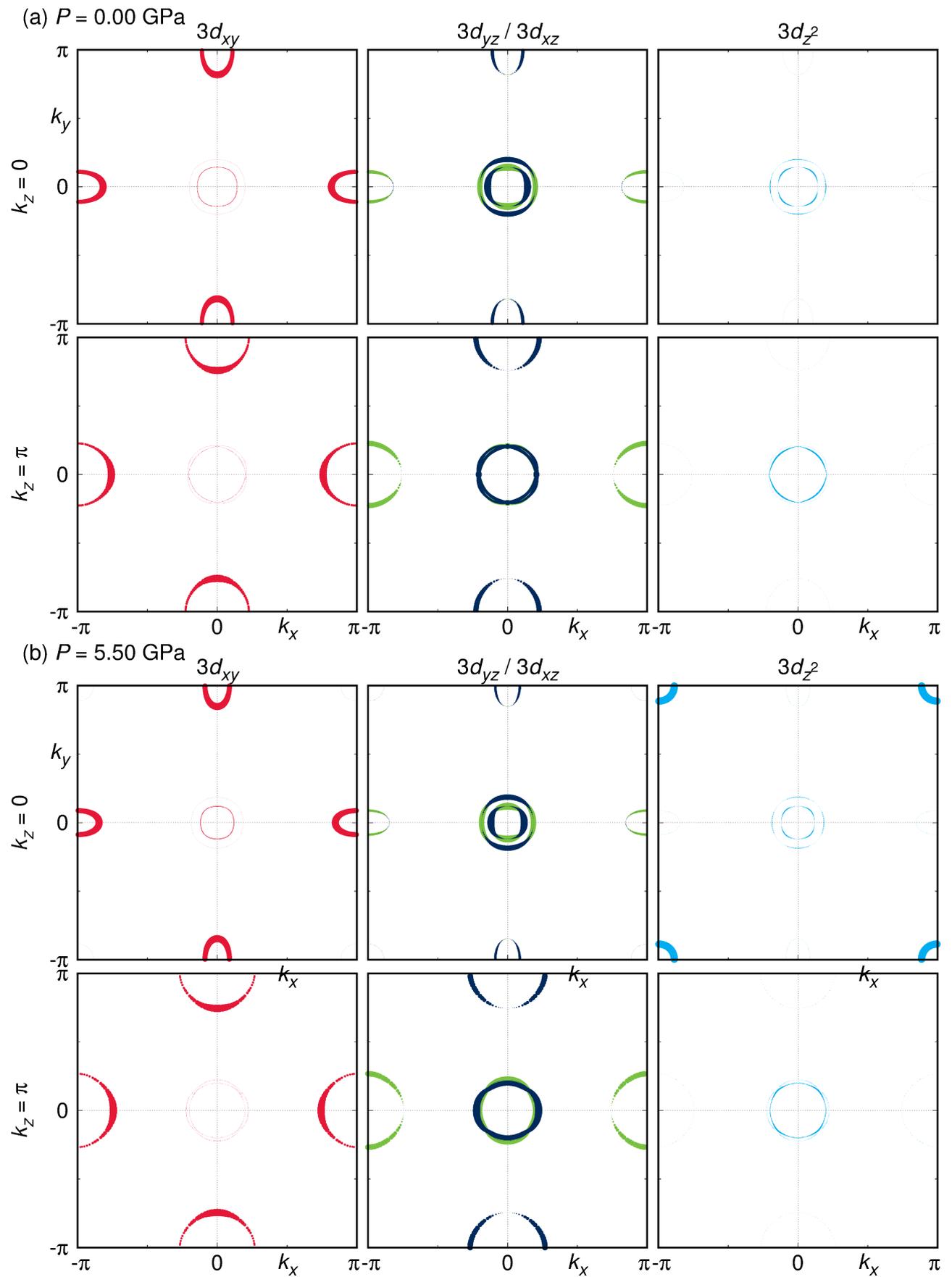

FIG. S5. Fermi surfaces of FeS at two different pressures with orbital weights. They are calculated from the tight binding models which were unfolded to the one-iron Brillouin zone.

$k_z = 0$ and $k_z = \pi$ shows that the $P = 5.5$ GPa electronic structure is more three-dimensional than the ambient pressure electronic structure. Three-dimensional susceptibility and pairing calculations are essential for properly describing the pressure dependence of superconductivity in FeS.

### F. Off-diagonal components of the spin susceptibility

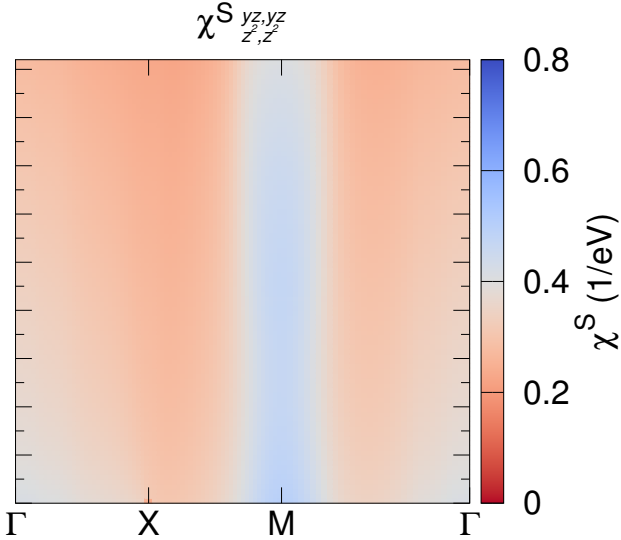

FIG. S6. Pressure dependence of an off-diagonal element of the susceptibility for tetragonal FeS.

Figure S6 shows a relevant off-diagonal component of the spin susceptibility $\chi^{S\,bb}_{aa}$ with $a = d_{z^2}$ and $b = d_{yz}$. The even more important off-diagonal component of $\chi^{S\,bb}_{aa}$ with $a = d_{z^2}$ and $b = d_{xy}$ is shown in the main text (Figure 2 (d)). $\chi^{S\,bb}_{aa}$ with $a = d_{z^2}$ and $b = d_{yz}$ has a weak maximum near $\mathbf{q} = (\pi, \pi)$, due to some nesting between $d_{yz}/d_{xz}$ character hole pockets around $\Gamma$ and the $d_{z^2}$ character hole pocket around $M$.

### G. Solution of the gap equation with intra-orbital interaction only

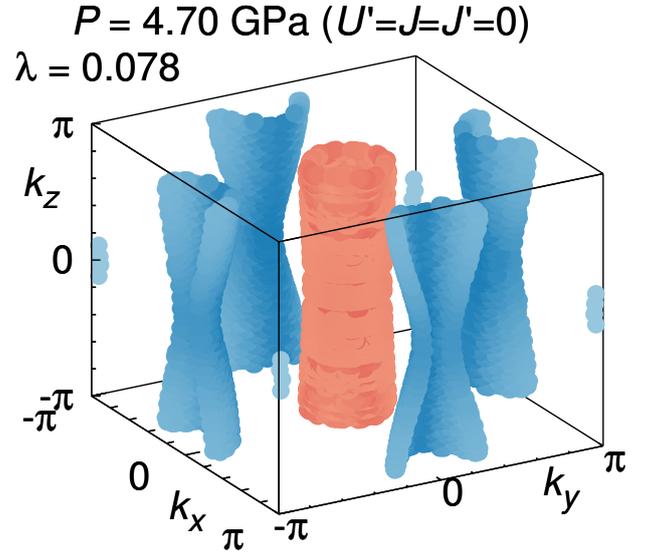

FIG. S7. Hypothetical leading gap function for FeS at $P = 4.7$ GPa for $U = 1.9$ eV and $U' = J = J' = 0$. The three-dimensional Fermi surface is plotted in the one-iron Brillouin zone.

In Figure S7, we demonstrate the effect of inter-orbital interaction terms by determining the solution of the gap equation at $U' = J = J' = 0$. The solution at $P = 4.7$ GPa (and higher pressures) is a simple nodeless sign-changing $s$ wave instead of the more complicated $s_\pm$ solution shown in Figure 3 (c) of the main text.

It arises from the $\mathbf{q} = (\pi, 0), (0, \pi)$ nesting of the $d_{yz}/d_{xz}$ orbitals. Note also the remaining orbital weight of $d_{xy}$ and $d_{z^2}$ around $\Gamma$ in Figs. 3(a) and (c) of the main text, which enables these orbitals to participate in a $s_\pm$ solution.


%\clearpage
%\includepdf[pages=7]{Supplemental_Material.pdf}
%\clearpage
%\includepdf[pages=8]{Supplemental_Material.pdf}


\begin{thebibliography}{99}

% References for introduction

\bibitem{Hsu2008} F.-C. Hsu, J.-Y. Luo, K.-W. Yeh, T.-K. Chen,
  T.-W. Huang, P. M. Wu, Y.-C. Lee, Y.-L. Huang, Y.-Y. Chu, and D.-C.
  Yan, and M.-K. Wu, {\it Superconductivity in the PbO-type structure
  $\alpha$-FeSe}, Proc. Natl. Acad. Sci. U.S.A. {\bf 105}, 14262
  (2008).

\bibitem{Kamihara2008} Y. Kamihara, T. Watanabe, M. Hirano, and
  H. Hosono, {\it Iron-Based Layered Superconductor La[O$_{1-x}$F$_x$]FeAs
  ($x = 0.05$-$0.12$) with $T_{\rm c} = 26$ K}, J.  Am. Chem. Soc. {\bf
    130}, 3296 (2008).

\bibitem{Boehmer2018} A. E B{\"o}hmer and A. Kreisel, {\it Nematicity,
  magnetism and superconductivity in FeSe}, J. Phys.: Condens. Matter
  {\bf 30}, 023001 (2018).

\bibitem{Glasbrenner2015} J. K. Glasbrenner, I. I. Mazin,
  H. O. Jeschke, P. J. Hirschfeld, R.M. Fernandes, and R. Valent\'{i},
  {\it Effect of magnetic frustration on nematicity and superconductivity
  in iron chalcogenides}, Nat. Phys. {\bf 11}, 953 (2015).

\bibitem{Coldea2018} A. Coldea and M. D. Watson, {\it The key ingredients
  of the electronic structure of FeSe}, Annu. Rev. Condens. Matter
  Phys. {\bf 9}, 125 (2018).

\bibitem{Lai2015} X. Lai, H. Zhang, Y. Wang, X. Wang, X. Zhang,
  J. Lin, and F. Huang, {\it Observation of Superconductivity in Tetragonal
  FeS}, J. Am. Chem. Soc. {\bf 137}, 10148 (2015).

\bibitem{Watson2015} M. D. Watson, T. K. Kim, A. A. Haghighirad,
  S. F. Blake, N. R. Davies, M. Hoesch, T. Wolf, A. I. Coldea,
  {\it Suppression of orbital ordering by chemical pressure in
  FeSe$_{1-x}$S$_x$}, Phys. Rev. B {\bf 92}, 121108 (2015).

\bibitem{Man2017} H. Man, J. Guo, R. Zhang, R. Sch{\"o}nemann, Z. Yin,
  M. Fu, M. B. Stone, Q. Huang, Y Song, W. Wang, D. J. Singh,
  F. Lochner, T. Hickel, I. Eremin, L. Harriger, J. W. Lynn,
  C. Broholm, L. Balicas, Q. Si and P. Dai, {\it Spin excitations and the
  Fermi surface of superconducting FeS}, NPJ Quant. Mater. {\bf 2}, 14
  (2017).

\bibitem{Borg2016} C. K. H. Borg, X. Zhou, C. Eckberg, D. J. Campbell,
  S. R. Saha, J. Paglione, E. E. Rodriguez, {\it Strong anisotropy in
  nearly ideal tetrahedral superconducting FeS single crystals},
  Phys. Rev. B {\bf 93}, 094522 (2016).

\bibitem{Hosoi2016} S. Hosoi, K. Matsuura, K. Ishida, H. Wang,
  Y. Mizukami, T. Watashige, S. Kasahara, Y. Matsuda, and
  T. Shibauchi, {\it Nematic quantum critical point without magnetism in
  {\ses} superconductors}, Proc. Natl. Acad. Sci. {\bf 113}, 8139
  (2016).

\bibitem{Matsuura2017} K. Matsuura, Y. Mizukami, Y. Arai, Y. Sugimura,
  N. Maejima, A. Machida, T. Watanuki, T. Fukuda, T. Yajima, Z. Hiroi,
  K. Y. Yip, Y. C. Chan, Q. Niu, S. Hosoi, K. Ishida, K. Mukasa,
  T. Watashige, S. Kasahara, J.-G. Cheng, S. K. Goh, Y. Matsuda,
  Y. Uwatoko, T. Shibauchi, {\it Maximizing {\tc} by tuning nematicity and
  magnetism in {\ses} superconductors}, Nat. Commun. {\bf 8}, 1143
  (2017).

\bibitem{Sato2018} Y. Sato, S. Kasahara, T. Taniguchi, X. Xing,
  Y. Kasahara, Y. Tokiwa, Y. Yamakawa, H. Kontani, T. Shibauchi, and
  Y. Matsuda, {\it Abrupt change of the superconducting gap structure at
  the nematic critical point in {\ses}}, Proc. Natl. Acad. Sci. {\bf
    115}, 1227 (2018).
 
\bibitem{Hanaguri2018} T. Hanaguri, K. Iwaya, Y. Kohsaka, T. Machida,
  T. Watashige, S. Kasahara, T. Shibauchi, Y. Matsuda, {\it Two distinct
  superconducting pairing states divided by the nematic end point in
  {\ses}}, Sci. Adv. {\bf 4}, eaar6419 (2018).

\bibitem{Kuhn2017} S.J. Kuhn, M. K. Kidder, D. S. Parker, C. dela
  Cruz, M. A. McGuire, W. M. Chance, L. Li, L. Debeer-Schmitt,
  J. Ermentrout, K. C. Littrell, M. R. Eskildsen, A. S. Sefat,
  {\it Structure and property correlations in FeS}, Physica C {\bf 534}, 29
  (2017).

\bibitem{XYang2016} X. Yang, Z. Du, G. Du, Q. Gu, H. Lin, D. Fang,
  H. Yang, X. Zhu, H.-H. Wen, {\it Strong-coupling superconductivity
  revealed by scanning tunneling microscope in tetragonal FeS},
  Phys. Rev. B {\bf 94}, 024521 (2016).

\bibitem{Lin2016} H. Lin, Y. Li, Q. Deng, J. Xing, J. Liu, X. Zhu,
  H. Yang and H.-H. Wen, {\it Multiband superconductivity and large
  anisotropy in FeS crystals}, Phys. Rev. B {\bf 93}, 144505 (2016).

\bibitem{Xing2016} J. Xing, H. Lin, Y. Li, S. Li, X. Zhu, H. Yang,
  H.-H. Wen, {\it Nodal superconducting gap in tetragonal FeS}, Phys. Rev. B
  {\bf 93}, 104520 (2016).

\bibitem{Ying2016} T. P. Ying, X. F. Lai, X. C. Hong, Y. Xu, L. P. He,
  J. Zhang, M. X. Wang, Y. J. Yu, F. Q. Huang, and S. Y. Li, {\it Nodal
  superconductivity in FeS: Evidence from quasiparticle heat
  transport}, Phys. Rev. B {\bf 94}, 100504(R) (2016).

\bibitem{Holenstein2016} S. Holenstein, U. Pachmayr, Z. Guguchia,
  S. Kamusella, R. Khasanov, A. Amato, C. Baines, H.-H. Klauss,
  E. Morenzoni, D. Johrendt, and H. Luetkens, {\it Coexistence of
  low-moment magnetism and superconductivity in tetragonal FeS and
  suppression of {\tc} under pressure}, Phys. Rev. B {\bf 93}, 140506(R)
  (2016).

\bibitem{Kirschner2016} F. K. K. Kirschner, F. Lang, C. V. Topping,
  P. J. Baker, F. L. Pratt, S. E. Wright, D. N. Woodruff,
  S. J. Clarke, and S. J. Blundell, {\it Robustness of superconductivity to
  competing magnetic phases in tetragonal FeS}, Phys. Rev. B {\bf 94},
  134509 (2016).

\bibitem{YYang2016} Y. Yang, W.-S. Wang, H.-Y. Lu, Y.-Y. Xiang, and
  Q.-H. Wang, {\it Electronic structure and {\dw}-wave superconductivity in
  FeS}, Phys. Rev. B {\bf 93}, 104514 (2016).

\bibitem{Lai2016} X. Lai, Y. Liu, X. L{\"u}, S. Zhang, K. Bu, C. Jin,
  H. Zhang, J. Lin and F. Huang, {\it Suppression of superconductivity and
  structural phase transitions under pressure in tetragonal FeS},
  Sci. Rep. {\bf 6}, 31077 (2016).

\bibitem{Sun2012} L. Sun, X.-J. Chen, J. Guo, P. Gao, Q.-Z. Huang,
  H. Wang, M. Fang, X. Chen, G. Chen, Q. Wu, C. Zhang, D. Gu, X. Dong,
  L. Wang, K. Yang, A. Li, X. Dai, H.-k. Mao and Z. Zhao, {\it Re-emerging
  superconductivity at 48 kelvin in iron chalcogenides}, Nature {\bf
    483}, 67 (2012).

\bibitem{Izumi2015} M. Izumi, L. Zheng, Y. Sakai, H. Goto, M. Sakata,
  Y. Nakamoto, H. L. T. Nguyen, T. Kagayama, K. Shimizu, S. Araki,
  T. C. Kobayashi, T. Kambe, D. Gu, J. Guo, J. Liu, Y. Li, L. Sun,
  K. Prassides and Y. Kubozono, {\it Emergence of double-dome
  superconductivity in ammoniated metal-doped FeSe}, Sci. Rep. {\bf 5},
  9477 (2015).

\bibitem{Sun2018} J.P. Sun, P. Shahi, H. X. Zhou, Y. L. Huang,
  K. Y. Chen, B. S. Wang, S. L. Ni, N. N. Li, K. Zhang, W. G. Yang,
  Y. Uwatoko, G. Xing, J. Sun, D. J. Singh, K. Jin, F. Zhou,
  G. M. Zhang, X. L. Dong, Z. X. Zhao and J.-G. Cheng, {\it Reemergence of
  high-Tc superconductivity in the (Li$_{1-x}$Fe$_x$)OHFe$_{1-y}$Se under high
  pressure}, Nat. Commun. {\bf 9}, 380 (2018).

\bibitem{Das2016} T. Das and C. Panagopoulos, {\it Two types of
  superconducting domes in unconventional superconductors}, New
  J. Phys. {\bf 18}, 103033 (2016).

\bibitem{Lennie1995} A. R. Lennie, S. A. T. Redfern, P. F. Schofield,
  D. J. Vaughan, {\it Synthesis and Rietveld crystal structure refinement
  of mackinawite}, tetragonal FeS, Mineral. Mag. {\bf 59}, 677 (1995).

\bibitem{Zhang2017} J.  Zhang, F.-L. Liu, T.-P. Ying, N.-N. Li, Y. Xu,
  L.-P. He, X.-C. Hong, Y.-J. Yu, M.-X. Wang, J. Shen, W.-G. Yang and
  S.-Y. Li, {\it Observation of two superconducting domes under pressure in
  tetragonal FeS}, NPJ Quant. Mater. {\bf 2}, 49 (2017).

\bibitem{Ehm2008} L. Ehm, F. M. Michel, S. M. Antao, C. D. Martin,
  P. L. Lee, S. D. Shastri, P. J. Chupasc and J. B. Parise, {\it Structural
  changes in nanocrystalline mackinawite (FeS) at high pressure},
  J. Appl. Crystallogr. {\bf 42}, 15 (2009).

\bibitem{Koepernik1999} K. Koepernik and H. Eschrig, {\it Full-potential
  nonorthogonal local-orbital minimum-basis band-structure scheme},
  Phys. Rev. B \textbf{59}, 1743 (1999); http://www.FPLO.de

\bibitem{Perdew1996} J. P. Perdew, K. Burke, and
  M. Ernzerhof, {\it Generalized Gradient Approximation Made Simple},
  Phys. Rev. Lett. \textbf{77}, 3865 (1996).

\bibitem{Eschrig2009} H. Eschrig and K. Koepernik, {\it Tight-binding
  models for the iron-based superconductors}, Phys. Rev. B \textbf{80},
  104503 (2009).

\bibitem{Tomic2014} M. Tomi\'c, H. O. Jeschke, and R. Valent\'i,
  {\it Unfolding of electronic structure through induced representations of
  space groups: Application to Fe-based superconductors}, Phys. Rev. B
  \textbf{90}, 195121 (2014).

\bibitem{Graser2009} S. Graser, T. A. Maier, P. J. Hirschfeld, and
  D. J. Scalapino, {\it Near-degeneracy of several pairing channels in
  multiorbital models for the Fe pnictides}, New J. Phys. \textbf{11},
  025016 (2009).

\bibitem{Supplement} See the Supplementary information for details on
  the crystal structures and the spin fluctuation pairing
  calculations.

\bibitem{Guterding2015a} D. Guterding, H. O. Jeschke,
  P. J. Hirschfeld, and R. Valent\'i, {\it Unified picture of the doping
  dependence of superconducting transition temperatures in alkali
  metal/ammonia intercalated FeSe}, Phys. Rev. B {\bf 91}, 041112(R)
  (2015).

\bibitem{Guterding2017} D. Guterding, {\it Microscopic modelling of
    organic and iron-based superconductors}, PhD thesis,
  Goethe-Universit\"at Frankfurt am Main, Germany (2017).

\bibitem{Miao2017} J. Miao, X. H. Niu, D. F. Xu, Q. Yao, Q. Y. Chen,
  T. P. Ying, S. Y. Li, Y. F. Fang, J. C. Zhang, S. Ideta, K. Tanaka,
  B. P. Xie, D. L. Feng, and F. Chen, {\it Electronic structure of FeS},
  Phys. Rev. B {\bf 95}, 205127 (2017).

\bibitem{Reiss2017} P. Reiss, M. D. Watson, T. K. Kim,
  A. A. Haghighirad, D. N. Woodruff, M. Bruma, S. J. Clarke, and
  A. I. Coldea, {\it Suppression of electronic correlations by chemical
  pressure from FeSe to FeS}, Phys. Rev. B {\bf 96}, 121103(R) (2017).

\bibitem{Terashima2016} T. Terashima, N. Kikugawa, H. Lin, X. Zhu,
  H.-H. Wen, T. Nomoto, K. Suzuki, H. Ikeda and S. Uji, {\it Upper critical
  field and quantum oscillations in tetragonal superconducting FeS},
  Phys. Rev. B {\bf 94}, 100503(R) (2016).

\bibitem{Kuroki2009} K. Kuroki, H. Usui, S. Onari, R. Arita, and
  H. Aoki, {\it Pnictogen height as a possible switch between high-$T_c$
  nodeless and low-$T_c$ nodal pairings in the iron-based
  superconductors}, Phys. Rev. B {\bf 79}, 224511 (2009).


\end{thebibliography}
\end{document}